\documentstyle[prd,aps,twocolumn,psfig]{revtex}

\begin{document}
\twocolumn[\hsize\textwidth\columnwidth\hsize\csname
@twocolumnfalse\endcsname
\tighten
\draft
\title{Double Inflation in Supergravity and the Boomerang Observations}
\author{Toshiyuki Kanazawa} \address{Department of Physics, University
  of Tokyo, Tokyo, 113-0033, Japan} 
\author{Masahiro Kawasaki} \address{Research Center for the Early
  Universe (RESCEU), University of Tokyo, Tokyo, 113-0033, Japan}
\author{Naoshi Sugiyama} \address{Division of Theoretical
  Astrophysics, National Astronomical Observatory, Tokyo, 181-8588,
  Japan}
\author{T. Yanagida} \address{Department of Physics and RESCEU,
  University of Tokyo, Tokyo, 113-0033, Japan}
\date{\today}
\maketitle

\begin{abstract}

One of the biggest mystery of the recent observation of cosmic microwave
background anisotropies by the boomerang team is insignificance of the
second acoustic peak in the angular power spectrum.  It is very
difficult to explain such a low amplitude without assuming the higher baryon
density than predicted by the Big Bang Nucleosynthesis (BBN).  Employing
the double inflation model in supergravity, we show that the density
fluctuations produced by this inflation model can produce a sufficient
low second acoustic peak with the standard value of the baryon density
from BBN.  It is shown that these density fluctuations are also
consistent with the observations of cluster abundances and galaxy
distributions.

\end{abstract}
]

\section{Introduction}

After the discovery by COBE/DMR~\cite{smoot}, anisotropies of Cosmic
Microwave Background radiation (CMB) become one of the most important
targets of modern cosmology.  Theoretical works reveal that they
contain rich information, i.e., geometry of the universe, the baryon
density, the total matter density, the Hubble parameter, thermal
history and so on~\cite{HSS}.  However, it turns out that the angular
resolution of COBE/DMR was too crude to obtain above information.
Precise measurements of so-called acoustic peaks in the angular power
spectrum on arcminute scales are necessary to determine these
cosmological parameters.

There are many attempts to observe CMB anisotropies on arcminute
scales after COBE/DMR discovery\cite{max}.  Recently, the boomerang
team has reported very clear evidence of the first acoustic peak in
the angular power spectrum of CMB anisotropies~\cite{boomerang}.  The
location of the peak suggests the flatness of the universe.  However,
there remain some mysteries in their result.  One of them is a
relatively low second acoustic peak if there exists. The most natural
way to explain such a low peak is to increase the value of $\Omega_B
h^2$, where $\Omega_B$ is the baryon density parameter and $h$ is the
non-dimensional Hubble constant normalized by $100\rm km/s/Mpc$, since
increasing $\Omega_B h^2$ boosts only odd number peaks~\cite{HSS}. In
order to fit the data, however, $\Omega_B h^2$ needs to be larger than
the value constrained by Big Bang Nucleosynthesis
(BBN)~\cite{boomana}.

There are several possibilities to explain the low amplitude of the
second acoustic peak.  Among them, a tilted initial power spectrum,
increasing the diffusion length~\cite{wsp}, and degenerated
neutrinos~\cite{dneu} can be considered.  However, it seems that none of
them provides a small enough second-first peak height ratio without
increasing $\Omega_B h^2$. Here we propose double inflation which
breaks a coherent feature of the initial power spectrum as a candidate
for solving this mystery.

Recently we studied a double inflation model with hybrid and new
inflations and its cosmological implication~\cite{KKSY-LSS}.  It was
found that both inflations could produce cosmologically relevant
density fluctuations if the total $e$-fold number of new inflation is
small enough.  In this case, there appears a breaking scale in the
density power spectrum which corresponds to the horizon scale of the
transit epoch from hybrid to new inflation.  The fluctuations on
scales larger (smaller) than the breaking are produced by hybrid (new)
inflation.  Assuming the 'standard' cold dark matter model with
$\Omega_0=1$, where $\Omega_0$ is the density parameter of a matter
component, we can fit both cluster abundances and galaxy distributions
with the COBE/DMR normalization when we set the amplitude of
perturbations produced by new inflation is smaller than the one by
hybrid inflation.  Accordingly we generally obtain smaller acoustic
peaks in the CMB angular power spectrum since the peaks are controlled
by new inflation and the Sachs-Wolfe tail on larger scales whose
amplitude is fixed by the COBE normalization is determined by hybrid
inflation.

Therefore, we may be able to explain cluster abundances, galaxy
distributions and the low second acoustic peak at once by introducing
double inflation if the braking scale is in between first and second
peaks by chance.  In this paper, we first explain the double inflation
model and compare the resultant density power spectrum and CMB angular
power spectrum with observational data.
We take $\Omega_{0}+\lambda_{0}=1$, i.e., a spatially flat universe.
Here $\lambda_{0}$ is the density parameter of a cosmological
constant. We also take $\Omega_{B}h^{2}=0.02$ which is the best value
from BBN analysis~\cite{Omegab}.

\section{Double inflation model}

We adopt a double inflation model proposed in
Ref.\cite{dynamical-tuning}.
Here we briefly describe the model in order to show
the relation between model parameters and the power spectrum of the
density fluctuations.
The model consists of two inflationary stages; the first one is called
preinflation. Here we employ hybrid inflation~\cite{hybrid} (see also
Ref.\cite{hybrid2}) as preinflation. We also assume that the second
inflationary stage is realized by a new inflation
model~\cite{Izawa-New-inflation} and its $e$-fold number is smaller
than $\sim 60$.  Thus, the density fluctuations on large scales are
produced during preinflation and their amplitudes should be normalized
to the COBE data~\cite{COBE}.  On the other hand, new inflation
produces fluctuations on small scales.
Thus, this power spectrum has a break on the scale corresponding to a
turning epoch from preinflation to new inflation.  As for the detailed
argument of dynamics of our model, see
Refs.~\cite{KKSY-LSS,KSY-PBH,KY-PBH}.

\subsection{First inflationary stage}

First, let us briefly discuss a hybrid inflation model~\cite{hybrid}.
The hybrid inflation model contains two kinds of superfields: one is
$S(x,\theta)$ and the others are a pair of $\Psi(x,\theta)$ and
$\bar{\Psi}(x,\theta)$. Here $\theta$ is the Grassmann number denoting
superspace. The model is based on the U$(1)_R$ symmetry under which
$S(\theta) \rightarrow e^{2i\alpha} S(\theta e^{-i\alpha})$ and
$\Psi(\theta) \bar{\Psi}(\theta) \rightarrow \Psi(\theta e^{-i\alpha})
\bar{\Psi}(\theta e^{-i\alpha})$. The superpotential is given
by~\cite{hybrid}
\begin{equation}
\label{superpot-pre}
    W(S,\Psi,\bar{\Psi}) = -\mu^{2} S + \lambda S \bar{\Psi}\Psi.
\end{equation}
The $R$-invariant K\"ahler potential is given by
\begin{equation}
\label{kahlerpot-pre}
    K(S,\Psi,\bar{\Psi}) = |S|^{2} + |\Psi|^{2} + |\bar{\Psi}|^{2}
    + \cdots ,
\end{equation}
where the ellipsis denotes higher-order terms which we neglect in the
present analysis for simplicity. We gauge the U$(1)$ phase
rotation:$\Psi \rightarrow e^{i\delta}\Psi$ and $\bar\Psi \rightarrow
e^{-i\delta}\bar\Psi$. To satisfy the $D$-term flatness condition we
take always $\Psi = \bar\Psi$ in our analysis. 
For $|S|>|S_{c}|=\mu/\sqrt{\kappa}$, the effective potential $V$ has a
minimum at $\Psi=\bar{\Psi}=0$. That is, for $|S|>|S_{c}|$, the energy
density is dominated by the false vacuum energy density $\mu^{4}$ and
inflation takes place.
We identify the inflaton field $\sigma/ \sqrt{2}$ with the real part
of the field $S$.

We define $N_{\rm COBE}$ as the $e$-fold number corresponding to the
COBE scale and the COBE normalization leads to a condition for the
inflaton potential,
\begin{equation}
  \label{eq:COBE-cond}
  \left| \frac{V^{3/2}}{V'} \right|_{N_{\rm COBE}} \simeq \frac{4\pi
    \mu^{2}\sqrt{N_{\rm COBE}}}{\lambda} \simeq 5.3\times 10^{-4},
\end{equation}
where $V$ is the inflaton potential obtained from
Eqs.(\ref{superpot-pre}) and (\ref{kahlerpot-pre}) including one-loop
corrections. In the hybrid inflation model, density fluctuations are
almost scale invariant, $n_{\rm pre}\simeq 1$, where $n_{\rm pre}$ is
a spectral index for a power spectrum of density fluctuations.

\subsection{Second inflationary stage}

Now, we consider a new inflation model.  We adopt an inflation model
proposed in Ref.~\cite{Izawa-New-inflation}.  The inflaton superfield
$\phi(x, \theta)$ is assumed to have an $R$ charge $2/(n+1)$ and
U$(1)_{R}$ is dynamically broken down to a discrete $Z_{2nR}$ at a
scale $v$, which generates an effective
superpotential~\cite{Izawa-New-inflation,dynamical-tuning},
\begin{equation}
        W(\phi) = v^{2}\phi - \frac{g}{n+1}\phi^{n+1}.
        \label{sup-pot2}
\end{equation}
The $R$-invariant effective K\"ahler potential is given by
\begin{equation}
    \label{new-kpot}
    K(\phi) = |\phi|^2 +\frac{\kappa}{4}|\phi|^4 + \cdots ,
\end{equation}
where $\kappa$ is a constant of order $1$. Hereafter we take $n=4$ and
$g=1$ for simplicity.

An important point on the density fluctuations produced by new
inflation is that it results in a tilted spectrum with spectral index
$n_{\rm new}$ given by
\begin{equation}
    \label{eq:new-index}
    n_{\rm new} \simeq 1 - 2 \kappa.
\end{equation}

\subsection{Initial value and fluctuations of the inflaton $\varphi$}

The crucial point observed in Ref.~\cite{dynamical-tuning} is that
preinflation sets dynamically the initial condition for new inflation.
We identify the inflaton field $\varphi(x)/\sqrt{2}$ with the real
part of the field $\phi(x)$. Then, the value of $\varphi$ at the
beginning of new inflation is given by~\cite{KY-PBH}
\begin{equation}
  \varphi_{b} \simeq \frac{\sqrt{2}}{\sqrt{\lambda}}v \left(
    \frac{v}{\mu} \right)^{2}.
\end{equation}
Therefore, in our model, $\varphi_{b}$, the $e$-fold number of the
second inflation ($N_{\rm new}$), and $N_{\rm COBE}$ are determined by
only model parameters. On the contrary, in the other double
inflation models $\varphi_{b}$ should be put by hand. In our model we
have four model parameters $(\mu, \lambda, v, \kappa)$ among which
$\mu$ is expressed by the other parameters with use of
Eq.~(\ref{eq:COBE-cond}).  Thus there are three free parameters.

\subsection{Numerical results}

We estimate density fluctuations in double inflation by calculating
the evolution of $\varphi$ and $\sigma$ numerically. For given
parameters $\kappa$ and $\lambda$, we obtain the breaking scale $k_b$ and
the amplitude of the density fluctuations $\delta_{b}$ produced at 
the beginning of
new inflation. Here, $k_{b}^{-1}$ is the comoving
breaking scale corresponding to the Hubble radius at the beginning of
new inflation. We can understand the qualitative dependence of $(k_b,
\delta_{b})$ on $(\kappa, \lambda)$ as follows: When $\kappa$ is
large, the slope of the potential for new inflation is too steep, and
new inflation cannot last for a long time.  Therefore, the break
occurs at smaller scales. In fact, we can express $k_{b}$ as
\begin{equation}
  k_{b}
  \sim \frac{1}{3000}h {\rm Mpc}^{-1} \exp \left[50-
    \frac{1}{\kappa}\ln \left( \sqrt{\frac{\lambda(1-\kappa)}{12}}
      \frac{\mu^{2}}{v^{2}} \right) \right].
\end{equation}
As for $\delta_{b}$, we can see from Eq.(\ref{eq:COBE-cond}) that as
$\lambda$ becomes larger, $\mu$ also must become larger. In addition, we can
show that
\begin{equation}
  \delta_{b}\equiv \left( \frac{\delta \rho}{\rho} \right)_{{\rm new},
    k_{b}} \simeq \frac{1}{5\sqrt{6}\pi} \frac{\sqrt{\lambda}
    \mu^2}{\kappa v},
\end{equation}
for a given $v$ (see Ref.~\cite{KKSY-LSS}). Thus, we have larger
$\delta_{b}$ for larger $\lambda$.

In our previous work~\cite{KKSY-LSS}, we have shown that if
$\lambda\sim {\cal O}(10^{-4}-10^{-3})\quad {\rm and}\quad 0.1
\lesssim \kappa \lesssim 0.2$ , $k_b$ is at a cosmological scale
($10^{-3}h{\rm Mpc^{-1}} \lesssim k_b \lesssim 1 h{\rm Mpc^{-1}}$),
and density fluctuations produced during new inflation are not too far
from those of preinflation ($0.1 \lesssim {\cal R}\equiv P_{\rm
  new}/P_{\rm pre} \lesssim 10$).  Here $P_{\rm new}$ and $P_{\rm
  pre}$ refer to the amplitude of the power spectrum of the density
fluctuations at $k_b$, produced by new inflation and preinflation,
respectively:
\begin{equation}
P(k) = \left\{
\begin{array}{l}
  P_{\rm pre} \left( \frac{k}{k_b}\right)^1 T^2(k)\quad (k<k_b),\\ 
  P_{\rm new} \left( \frac{k}{k_b}\right)^{n_{\rm new}} T^2(k)\quad
  (k>k_b),
\end{array}
\right.
\end{equation}
where $T(k)$ is a matter transfer function.

\section{Comparison with observations}

\subsection{Second acoustic peak}

The spectral index of new inflation $n_{\rm new}$ is $n_{\rm new}
\simeq 1 - 2 \kappa < 1$ [see Eq.(\ref{eq:new-index})]. Also, the
amplitude of the density fluctuations on smaller scales, which are
produced during new inflation, can be smaller than that on larger
scales, which is normalized to the COBE/DMR data.  Thus, in our double
inflation model, if the breaking scale $k_{b}$ is in between the first and
the second peaks of the CMB angular power spectrum, there is a
possibility to explain the lower second acoustic peak of the boomerang
results.

Here we take $n_{\rm new} \simeq 0.8\ (\kappa\simeq 0.1)$ as an
example. Since the location of the first acoustic peak (the multipole
moment $\ell\sim 200$) corresponds to the comoving wave number $k\sim
{\cal O}(10^{-2})h$Mpc$^{-1}$, we have searched parameter sets within
the parameter range of $0.001 h{\rm Mpc}^{-1} \lesssim k_{b}\lesssim
0.04 h{\rm Mpc}^{-1}$, and $0.7\lesssim {\cal R} \lesssim 1$, which
can produce a lower second acoustic peak of the CMB angular power
spectrum.  

\begin{figure}[htbp]
  \centerline{\psfig{figure=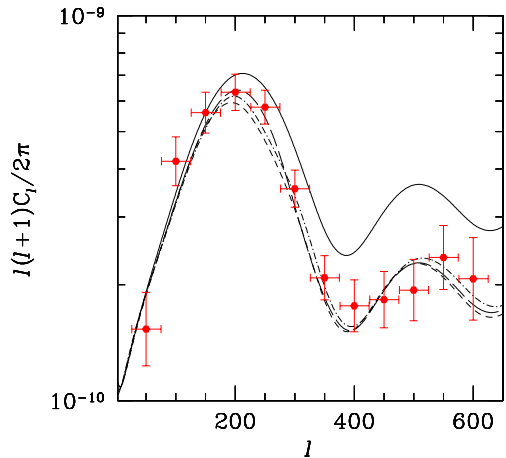,width=7cm}}
    \caption{The CMB angular power spectra for four parameter sets.
      The symbols with error bars are the boomerang data. The short
      dash line, long dash line, and dot-short dash line correspond to
      model (a) to (c) in Table~\protect\ref{table}, respectively. For
      comparison, we also plot the Lambda CDM model (model (d) in
      Table~\protect\ref{table}) in a solid line. }
    \label{fig:C_ell_samples}
\end{figure}

In Table \ref{table}, samples of these parameters are listed. From the
recent observations of Type Ia supernovae, $\Omega_{0}$ is estimated
as $\Omega_{0}\lesssim 0.5$~\cite{SNIa}. Therefore, we take
$\Omega_{0}=0.4, 0.5,$ and $0.6$, for example.
Also, in Fig.\ref{fig:C_ell_samples}, we plot the angular power spectrum
of the CMB anisotropies for these parameters. From this
figure, we can see that our double inflation model has a parameter
region which can explain the lower second acoustic peak of the
boomerang observations as expected.

\subsection{Cluster abundances and galaxy distribution}

As we have seen in the previous subsection, our double inflation model
can explain the low second acoustic peak of the boomerang data.
However, we need to check whether it is consistent with other
observations.  In this subsection we compare the result of our double
inflation model with the observations of the cluster
abundances~\cite{Eke,Viana-Liddle} and galaxy
distributions~\cite{gals}.

Usually the constraint on the power spectrum from observations of the
cluster abundances is expressed in terms of $\sigma_{8}$, 
the specific mass fluctuations within a sphere of a radius of 
$8h^{-1}\rm Mpc$. Since the
power spectrum of the density fluctuations shows a break on the
cosmological scale in our double inflation model, we cannot simply
employ the value of $\sigma_8$ quoted by previous
works~\cite{Eke,Viana-Liddle}.  We need to calculate the cluster
abundances by using the Press-Schechter theory~\cite{Press-Schechter}.

When we determine the breaking scale $k_b$, the power spectrum ratio
${\cal R}\equiv P_{\rm new}/P_{\rm pre}$, and the spectral index for
new inflation $n_{\rm new}$, we can get the power spectrum up to
normalization {$A_{\rm cl}$}. Using this power spectrum we can
calculate the comoving abundance of the clusters as
\begin{equation}
  n(>M_{\rm min} ; A_{\rm cl}) = \int^{\infty}_{M_{\rm min}}
  \frac{dn(M)}{dM} dM,
\label{eq:abundance-PS}
\end{equation}
where mass distribution $dn/dM$ is obtained by the Press-Schechter
formula.

Many clusters of galaxies are observed with use of x-ray fluxes. Under
the assumption that clusters are hydrostatic, we can obtain the
mass-temperature relations as
\begin{equation}
  T_{\rm gas} = \frac{9.37\ {\rm keV}}{\beta (5X+3)} \left(
    \frac{M}{10^{15}h^{-1} M_\odot} \right)^{2/3} (1+z)
  \Omega_{0}^{1/3} \Delta_c^{1/3},
\label{eq:mass-temperature}
\end{equation}
where $\Delta_c$ is the ratio of the density of a cluster to the
background mean density at that redshift, $\beta$ is the ratio of
specific galaxy kinetic energy to specific gas thermal energy, and $X$
is the hydrogen mass fraction. Following Ref.~\cite{Eke}, we take
$X=0.76$, $\beta=1$. Also, $\Delta_c$ can be approximated as
$\Delta_{c} \simeq 18\pi^2\left[ 1+0.4093 \left(1/\Omega_{0} -
    1\right)^{0.9052} \right]$~\cite{Nakamura-Suto}.

The observed cluster abundance as a function of x-ray temperature can
be translated into a function of mass using
Eq.~(\ref{eq:mass-temperature}). Accumulating the observations, Henry
and Arnaud \cite{Henry-Arnaud} gave the fitting formula as
\begin{equation}
\left(\frac{dn(T)/dT}{h^{3}{\rm Mpc}^{-3}{\rm keV}^{-1}}\right) = 1.8
\left\{
  \begin{array}{l}
    +0.8\\
    -0.5
  \end{array}
\right\}
 \times 10^{-3} 
 \left( \frac{kT}{1{\rm keV}} \right)^{-4.7\pm 0.5}.
\label{eq:temperature-spectrum}
\end{equation}
Ref.~\cite{Henry-Arnaud} also gave a table of cluster observations
whose temperatures are larger than $2.5$ keV, which determines the
lower limit $M_{\rm min}$ from Eq.~(\ref{eq:mass-temperature}).
Therefore, by integrating Eq.~(\ref{eq:temperature-spectrum}) we obtain
\begin{equation}
  6.6\times 10^{-6}\ \lesssim\ n(>M_{\rm min})\ \lesssim\ 4.3\times
  10^{-5}.
\label{eq:abundance-obs}
\end{equation}
Matching these abundances, Eq.~(\ref{eq:abundance-PS}) calculated from
the Press-Schechter theory, and Eq.~(\ref{eq:abundance-obs}) inferred
from the x-ray cluster observations, we can determine the
normalization (amplitude) of power spectrum, $A_{\rm cl}$. Using this
normalization, we can obtain ``cluster abundance normalized''
$\sigma_8$, $\sigma_{\rm 8, cl}$, as
\begin{equation} 
  \left.  \sigma_{\rm 8, cl}^2 \equiv \int^{\infty}_{0}
    \frac{k^3}{2\pi^2} P(k; A_{\rm cl}) W^2(kr_0) \frac{dk}{k}
  \right|_{r_0=8h^{-1}{\rm Mpc}}.
\end{equation}
where $P(k; A)$ is a present matter density fluctuation power spectrum
with a normalization $A$, and $W(x)$ is a window function. Because of
errors in observations, we have some range for allowed $\sigma_{\rm 8,
  cl}$.

On the other hand, we can normalize the power spectrum by COBE
data~\cite{COBE,Bunn-White}.  Therefore, we have ``COBE normalized''
$\sigma_8$, $\sigma_{\rm 8, COBE}$ together with $\sigma_{\rm 8, cl}$.
Bunn and White~\cite{Bunn-White} estimates one standard deviation
error of COBE normalization to be $7\%$ which is much smaller than the
one of cluster normalization.  We assume that, therefore, if $\sigma_{\rm 8 ,
  COBE}$ lies in an allowed $\sigma_{\rm 8, cl}$ range, the parameter
region of $k_b$, ${\cal R}$, and $n_{\rm new}$ is consistent with the
cluster abundance observations.
As for the parameter sets for models (a) to (c) in Table~\ref{table},
we have confirmed that they all satisfy the cluster abundance
constraint.


We also have to investigate whether our parameter sets are consistent
with the observations of galaxy distributions.  There are many
observations which measure the density fluctuations from galaxy
distributions. Among them we use the data sets compiled by
Vogeley~\cite{Vogeley} from
Refs.~\cite{gals} in this paper.

Employing the COBE normalization, we can determine the power spectrum
with its overall amplitude if we fix the breaking scale $k_b$, the
power spectrum ratio ${\cal R}$, and the spectral index for new
inflation $n_{\rm new}$.  One might make direct comparison of this
power spectrum with above observations of galaxy distributions.
However, distribution of luminous objects such as galaxies could
differ from underlying mass distribution because of so-called bias.
There is even no guarantee that each observational sample has the same
bias factor.  Therefore, we only consider the shape of the power
spectrum here.  We change the overall amplitude of each set of
observations arbitrarily.  Thus, we estimate the goodness of fitting
by calculating $\chi^{2}$ of this power spectrum with fixing $k_b,
{\cal R}$, and $n_{\rm new}$.

\begin{figure}[htbp]
  \centerline{\psfig{figure=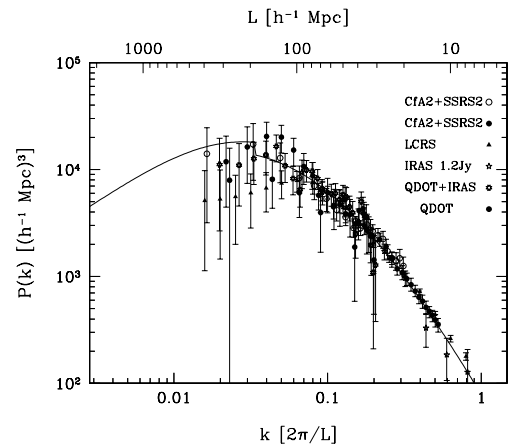,width=7cm}}
    \caption{The power spectrum of the density fluctuations for the
      model (b) in Table~\protect\ref{table}.}
    \label{fig:powerspectrum}
\end{figure}

For each parameter set we have chosen in the previous subsection, they
fit the observations of galaxy distributions well (reduced
$\chi^{2}_{\rm gal}\simeq 1$), except for model (a) [see
Table~\ref{table}].
In Fig.~\ref{fig:powerspectrum}, we plot the power spectrum of the
density fluctuations for the model (b) in Table~\ref{table} as an
example.

\section{Conclusions and discussions}

The boomerang team has reported that there is a low second acoustic peak
in the angular power spectrum of CMB anisotropies. Although there are
some explanations to this lower peak, they seem to need higher baryon
density than predicted by the Big Bang Nucleosynthesis. In this paper,
we have considered the double inflation model in supergravity, and
shown that the density fluctuations produced by this inflation model
can produce this low second acoustic peak.

Since the density fluctuations in our model has a nontrivial spectrum,
we have checked that it is consistent with the observations of the
cluster abundances and the galaxy distributions. We have found that
the fit to the data in our model is very good if we take
$\Omega_{0}\simeq 0.5$. We can conclude that the double inflation model can
account for the boomerang data without conflicting other observations.
In particular, we stress that our model does not require high baryon
density and hence is perfectly consistent with BBN.\\ 


T. K. is grateful to K. Sato for his continuous encouragement. A part
of this work is supported by Grant-in-Aid of the Ministry of Education
and by Grant-in-Aid, Priority Area ``Supersymmetry and Unified Theory
of Elementary Particles'' (\#707).

\begin{table}[h]
  \begin{center}
     \caption{The break parameters and reduced $\chi^{2}_{\rm gal}$.
       In the models (a) to (c), we employ $n_{\rm new}=0.8$.  For
   comparison, the cosmological constant dominated CDM model and 
      the 'standard' standard CDM model are shown in (d) and (e),
   respectively.  In all cases, 
       $\Omega_{B}h^{2}=0.02$.}
     \label{table}
     \begin{tabular}{|c||ccccc|}
       model & $\Omega_{0}$ & $h$ & $k_{b} [h{\rm
         Mpc}^{-1}]$ & ${\cal R}$ & reduced $\chi^{2}_{\rm gal}$\\ 
       \hline
       (a)& 0.4 & 0.7 & 0.025 & 0.78 & 1.65\\ 
       (b)& 0.5 & 0.65 & 0.034 &0.76 & 1.09\\ 
       (c)& 0.6 & 0.6 & 0.032 & 0.80 & 0.77\\ 
       (d)& 0.4 & 0.7 & $\cdots$ & 1.00 & 0.61\\ 
       (e)& 1.0 & 0.5 & $\cdots$ & 1.00 & 1.51
    \end{tabular}
  \end{center}
\end{table}

\end{document}